\documentclass[showpacs,floatfix,twocolumn]{revtex4-1}
\usepackage{graphicx,psfrag,amsmath,amssymb,amsfonts,latexsym,color,epsf,dcolumn,graphpap}
\usepackage{subfig}
\usepackage{float}
\usepackage{lipsum}
\usepackage{enumerate}

\usepackage{tikz-cd} 
\usepackage{tikz}
\usetikzlibrary{shapes.geometric, arrows}

\definecolor{red}{rgb}{1,0,0}
\definecolor{blue}{rgb}{0,0,1}
\definecolor{skyblue}{rgb}{0,0,.5}
\definecolor{green}{rgb}{0,1,0}
\definecolor{orange}{cmyk}{0,.4,1,0}

\newcommand{\fq}{\left(\frac{\hbar}{2 e}\right)^2}
\newcommand{\ELcav}{E_{l,\text{cav}}}
\begin{document}
\title{The quantum Otto cycle in a superconducting cavity in the non-adiabatic regime}

\author{Nicol\'as F.~Del Grosso$^1$ }
\author{Fernando C. Lombardo$^1$}
\author{Francisco D. Mazzitelli$^2$}
\author{Paula I.~Villar$^1$ }
\affiliation{$^1$ Departamento de F\'\i sica {\it Juan Jos\'e
 Giambiagi}, FCEyN UBA and IFIBA CONICET-UBA, Facultad de Ciencias Exactas y Naturales,
 Ciudad Universitaria, Pabell\' on I, 1428 Buenos Aires, Argentina }
\affiliation{$^2$ Centro At\'omico Bariloche and Instituto Balseiro,
Comisi\'on Nacional de Energ\'\i a At\'omica, 
R8402AGP Bariloche, Argentina}

\begin{abstract}
\noindent We analyze the efficiency of the quantum Otto cycle applied to a superconducting cavity. We consider its description in terms of a full quantum scalar field in a one-dimensional cavity with a time dependent boundary condition that can be externally controlled to perform and extract work unitarily from the system. We study the performance of this machine when acting as a heat engine as well as a refrigerator. It is shown that, in a non-adiabatic regime, the efficiency of the quantum cycle is affected by the dynamical Casimir effect, that induces a sort of quantum friction that diminishes the efficiency. We also find regions of parameters where the effect is so strong that the machine can no longer function as an engine since the work that would be produced is completely consumed by the quantum friction. However, this effect can be avoided for some particular temporal evolutions of the boundary conditions that do not change the occupation number of the modes in the cavity, leading to a highly improved efficiency.


\end{abstract}

\date{today}
\maketitle
\noindent 

\maketitle
\section{Introduction}\label{sec:intro}

As a consequence of the unceasing miniaturization of technological devices \cite{ion,circuitqed,nanoresonator},  there has been a growing interest in quantum thermodynamics in the last decade. This field captivates  two different but complementary features. On the one hand, it aims to obtain a rigorous derivation of the laws of thermodynamics from microscopic interactions at a quantum level. On the other hand, in a more applied aspect, it seeks to improve thermodynamic processes, such as the conversion of heat into mechanical work, using quantum phenomena without a classical analogue such as coherence \cite{coherence1,coherence2} or entanglement \cite{entanglement1,entanglement2}. The concept of information, and its intimate relationship with entropy and thermodynamics, plays a very important role in both aspects \cite{information} mentioned above.   

Likewise, the advent of new technologies pursuing  improvement in the experiments has attained the observation of phenomena in the laboratory that would have been unthinkable until recently. Conjectures and thermodynamic relationships can now be studied in multiple experimental implementations: from ion traps, through cold atoms in optical networks, to superconducting qubits and atom chips \cite{ottoion,optomechanicalengine,ottoqubit}. All these features are taken into consideration in a context of miniaturization of technology on the nano-scale. Hence, the question that naturally arises is to what extent the laws of thermodynamics and its phenomena are respected in the microscopic world. 

A fundamental role in the progress of quantum thermodynamics is played by  
small autonomous quantum thermal machines. These machines represent an ideal testing bench for studying quantum thermodynamics, as their size requires a quantum description for its evolution and they can provide work using thermal interactions with heat baths at different temperatures. In particular, it is very interesting to study how entanglement and coherence can enhance performance of these machines, for instance by achieving better cooling or extracting more work from given resources. Moreover, the investigation about the feasibility of experimental realizations of autonomous quantum thermal machines in mesoscopic systems, such as superconducting qubits and semiconductor quantum dots; or quantum simulations using standard circuit quantum electrodynamics architectures, has become relevant these days. Most of the research in this area has been conducted on qubits \cite{ottoqubit} or harmonic oscillators \cite{kosloff} subjected to different thermodynamic cycles. While in certain cases a quantum field in a cavity can be studied as a few modes that behave as harmonic oscillators, there are important circumstances under which this approximation fails. However, only a handful of papers have studied the effects arising from a full quantum field \cite{qftmachine,unruhengine1,unruhengine2,unruhengine3} and most of them as a bath and not a working medium.


In this work we shall study a thermal machine implemented with a superconducting circuit, consisting of a transmission line terminated by a superconducting quantum interference device (SQUID), which is subjected to a quantum Otto cycle. The machine is driven by an external magnetic field applied to the SQUID. For certain choices of the parameters of the circuit, the behavior of the machine is equivalent to that of a cavity with variable length, in which the quantum scalar field is the working medium. This provides an interesting connection with the systems usually considered to analyze the so called dynamical Casimir effect (DCE),  that will play an important role in what follows.

Broadly, the quantum Otto cycle involves a system, or working medium, ruled by a Hamiltonian $H_0$ to which four basic operations or strokes are applied in a cyclic fashion. Firstly, the system is put into contact with a cold bath at inverse temperature $\beta_A$;  leaving the system in a thermal state, with an internal energy $E_A$. Secondly, the system is isolated from the bath and subjected to a time dependent Hamiltonian; reaching a state with internal energy $E_B$. Thirdly, the system is put into contact with a hot bath at inverse temperature $\beta_C$; attaining once again a thermal state but this time of internal energy $E_C$. Finally, the system is once more isolated from the bath and subjected to a time dependent Hamiltonian that restores the original Hamiltonian $H_0$, leaving the system in the state with internal energy $E_D$.

As a consequence, in the third stroke we provide the machine with heat $$Q=E_C-E_B$$ and we  extract work between operations A and B given by $W_{AB}=E_A-E_B$ and between C and D given by $W_{CD}=E_C-E_D$. This amounts to a total work extracted of $$W=W_{AB}+W_{CD}.$$ If this work is positive, $W>0$, we say that the machine acts as a heat engine, which means it converts the heat from the thermal baths into useful work with an efficiency given by $$\eta=\frac{W}{Q}.$$ On the other hand, if the work and heat are negative, $W, Q < 0$, it is said that the machine acts as a refrigerator, which means it takes work and uses it to heat the hot bath and cool the cold one.


\section{ The System}

We shall consider a superconducting cavity of finite size (for example a waveguide ended with two SQUIDs). The electromagnetic field inside the cavity can be described by a quantum 
massless scalar field, the superconducting phase field $\Phi(x,t)$, where $x$ is the spatial coordinate along the cavity. The system can be therefore modeled by a massless scalar field in $1+1$ dimensions satisfying generalized Robin boundary conditions \cite{fosco}. The generalization involves not only time dependent parameters, but also the presence of the second time derivative $\partial_t^2 \Phi$. We will assume  the cavity has length $L_0$, it is decoupled from the input line  at $x=0$,  and has a SQUID at $x=L_0$. 
The cavity, which is assumed to have capacitance $c$ and inductance $l$ per unit length,   is described by the superconducting phase field Lagrangian
\begin{eqnarray}\label{lag}
L_{\rm cav} &=& \fq \frac{c}{2} \int_0^{L_0} d x \left((\partial_t \Phi)^2 - v^2 (\partial_x \Phi)^2 \right)\\ 
&+& \left[ \fq \frac{2 C_J}{2} \partial_t \Phi(L_0,t)^2 
 -  E_J \cos{f(t)} \Phi(L_0,t)^2
 \right]
 \,\nonumber ,
\end{eqnarray}
where  $v = 1/\sqrt{l c}$ is the field propagation velocity and $f(t)$ 
is the phase across the SQUID controlled by external magnetic flux. $E_J$ and $C_J$ denote the Josephson energy and capacitance, respectively. 
As anticipated, the description of the cavity involves the field $\Phi(x,t)$ for $0<x<L_0$ and the  additional degree of freedom $\Phi(L_0,t)$. The dynamical equations read
\begin{equation}
\partial_t^2 \Phi - v^2 \partial_x^2 \Phi = 0
\,,
\end{equation}
(in what follows we will set $v=1$), and  \cite{numerico1, numerico2}
\begin{eqnarray}\label{eqphid}
\frac{\hbar^2}{E_C} \partial_t^2 \Phi(L_0,t) &+& 2 E_J \cos{f(t)} \Phi(L_0,t)\\& + &\ELcav L_0 \partial_x \Phi(L_0,t)  = 0
\nonumber \,,
\end{eqnarray}
where $E_C = (2e)^2/(2 C_J)$ and $\ELcav = (\hbar/2e)^2 (1/l L_0)$. The equation above stems from the variation of the action with respect to $\Phi(L_0,t)$,
and can be considered as a generalized boundary condition for the field.   We could consider general boundary conditions also at $x=0$, but for the sake of simplicity  we will assume
that  $\Phi(0,t)=0$ (physically corresponding to the situation where the cavity is decoupled). Under a specific choice of the cavity and SQUID parameters, and also adjusting the external magnetic field across the SQUID at $x= L_0$, the second time derivative of the field becomes negligible and Eq.\eqref{eqphid} can be written as
\begin{align}
    0&= \Phi(L_0,t) + \frac{\ELcav L_0}{2 E_J \cos{f(t)}} \partial_x \Phi(L_0,t) \\
    &\approx \Phi(L_0+\frac{\ELcav L_0}{2 E_J \cos{f(t)}},t)
\end{align}
and the superconducting cavity behaves as a perfect cavity with a moving mirror at end, i.e. $L_0$ is a function of time, that we will denote $L_0 \equiv L(t)=L_0+\frac{\ELcav L_0}{2 E_J \cos{f(t)}}$. In this scenario, one may impose Dirichlet boundary conditions at both ends of the cavity, say $\Phi(0,t)=\Phi(L(t),t)=0$ \cite{louko, PRE, PRA2espejos}. In fact, it has been demonstrated that non stationary boundary conditions effects in a cavity can be implemented in a circuit QED system \cite{zeilinger,dyncasimir}.

For a static cavity, $L(t)=L_0$, these conditions determine  the eigenfrequencies, which are given by $n\pi/L_0$. We will denote 
the frequency spectrum by $\{\omega_n(L_0)\}_{n\in\mathbb{N}}$, 
since many of our results will be valid for more general boundary conditions. This allows us to expand the field in terms of bosonic operators $a_n$ as
\begin{equation}
    \Phi(x,t)=\sqrt{\frac{2}{L_0}}\sum_{n=1}^\infty \big[a_ne^{-i\omega_n(L_0)t}\sin(\omega_n(L_0)x)+h.c.\big].
\end{equation}
As a result, the Hamiltonian of the system for the static case is given by 
\begin{equation}
	H_0=H_{\text{free}}+E_{\rm SC}(L)
\end{equation} 
where $H_{\text{free}}=\sum_{k=1}^\infty \hbar \omega_{k} N_k$ with $ N_k=a_k^\dagger a_k$ and $\omega_{k}=\omega_k(L_0)$, while $E_{\rm SC}(L_0)=-\pi\hbar/(24L_0)$ is the energy corresponding to the static Casimir effect (vacuum energy inside the cavity for Dirichlet boundary conditions).

One could naively think that if the wall moves with a trajectory $L(t)$ the system would evolve according to the Hamiltonian given by $\sum_{k=1}^\infty \omega_{k}(L(t)) N_k+E_{\rm SC}(L(t))$ which happens to be the case for an adiabatic process or transformation (for example if the wall moves infinitely slow). However, a careful calculation shows that if the trajectory is given by 
\begin{equation}
\label{eq:L}
	L(t)=L_0+\delta L(t)=L_0[1-\epsilon\delta(t)] 
\end{equation} 
with $0=\delta(0)<\delta(t)<\delta(\tau)=1$ and $0<\epsilon\ll1$, then,
as shown in Appendix \ref{apen1}, the Hamiltonian of the quantum field is actually 
\begin{equation}
\label{eq:H}
	H(t)=H_0+H_1(t)
\end{equation}
where 
\begin{align}
\label{eq:H1}
	&H_1(t)=\hbar\sum_{k=1}^{\infty}\left[\omega_{k}^{\prime}\delta L(t) a_{k}^{\dagger}a_{k}+\frac{\omega_{k}^{\prime}}{2}\delta L(t)\left(a_{k}^{\dagger2}+a_{k}^{2}\right)\right]\nonumber\\
	&+\frac{\hbar}{2i}\sum_{k,j=1}^{\infty}\frac{\dot{\delta L}(t) }{L_{0}}g_{kj}\sqrt{\frac{\omega_{k}}{\omega_{j}}}\left(a_{k}a_{j}-a_{k}^{\dagger}a_{j}+a_{k}a_{j}^{\dagger}-a_{k}^{\dagger}a_{j}^{\dagger}\right).
\end{align}
The result is valid up to first order in $\epsilon$, and we have used the definitions $\omega_{k}^\prime=(d\omega_k/dL)(L_0)$ and 
\begin{equation}
 g_{kj}=-g_{jk}= L\int_{0}^{L}dx\left(\partial_{L}\psi_{k}\right)\psi_{j}.
\end{equation}

The key physics of the motion of the wall is captured by the different terms in $H_1$. There are essentially four distinct physical processes represented by each term. The first term is proportional to the original Hamiltonian and simply reflects the fact that when the wall is static in a different position the frequency spectrum varies. The second term proportional to $\left(a_{k}^{\dagger2}+a_{k}^{2}\right)$ is directly associated with DCE,  and uses the kinetic energy of the wall to spontaneously create pairs of photons inside the cavity in mode $k$, even if the initial state of the cavity is the vacuum (squeezing effect). Finally, the third  term contains two qualitatively different processes, the sum $\left(a_{k}a_{j}-a_{k}^{\dagger}a_{j}^{\dagger}\right)$ is also associated with the DCE and generates pairs of entangled photons in modes $k$ and $j$, while the sum $\left(-a_{k}^{\dagger}a_{j}+a_{k}a_{j}^{\dagger}\right)$ simply redistributes or scatters photons from mode $k$ to $j$.

\section{Otto cycle in a superconducting cavity }
\begin{figure}
	\begin{center}
		\includegraphics[scale=0.8]{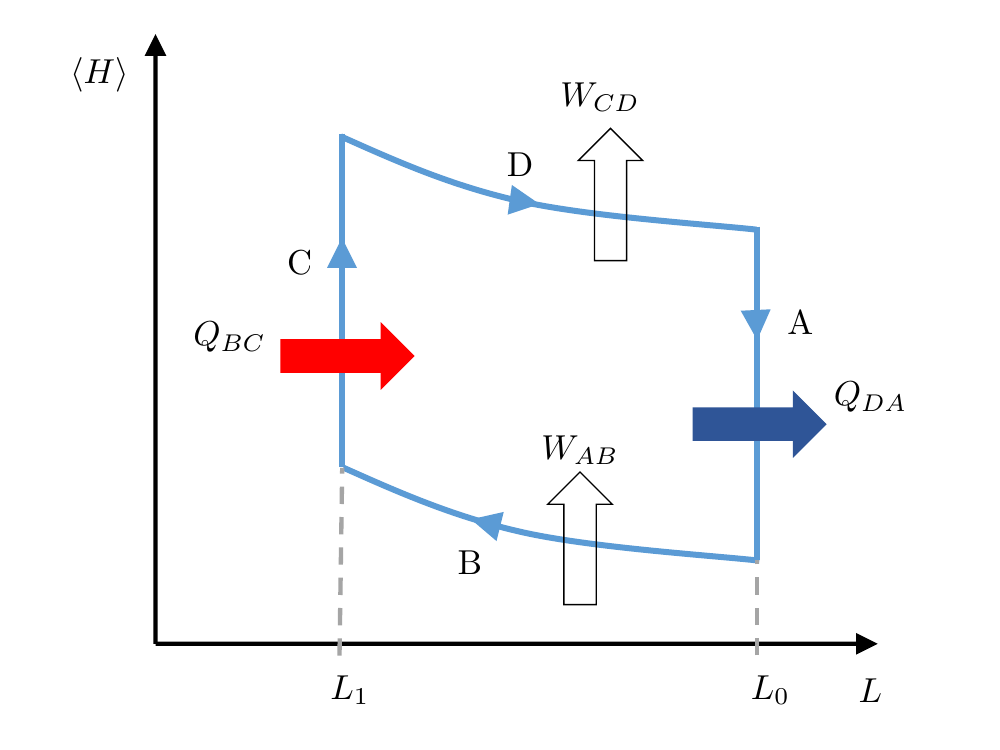}
		\caption{The four strokes of the Otto cycle in terms of the length of the cavity $L$ and the mean energy of the quantum field inside it.}
		\label{fig:esquema}
	\end{center}
\end{figure}

We will study the following implementation of the Otto cycle for a scalar quantum field in a cavity with a moving wall. The cycle is 
represented in Fig.\ref{fig:esquema} and can be described as: 

\begin{enumerate}[A.]
	\item First, the system is put into contact with a cold bath at inverse temperature $\beta_A$. This leaves the system in a thermal state 
	\begin{equation}
		\rho^{\beta_A}=\frac{\exp(-\beta_A H_{\text{free}})}{Z}=\prod_{k=1}^\infty \frac{e^{-\beta_A\omega_kN_k}}{Z_k}\, ,
	\end{equation}
	 with $$Z_k=\text{Tr}(e^{-\beta_A\hbar\omega_kN_k})=\frac{1}{1-e^{-\beta_A\hbar\omega_{k}}}$$ and internal energy 
	 \begin{align}
	 	E_A&=\text{Tr}({\rho^{\beta_A}H_{\text{free}}}/Z)+E_{\rm SC}(L_0)\nonumber\\
	 	&=\sum_{k=1}^\infty \frac{\hbar \omega_k}{e^{\beta_A\hbar\omega_{k}}-1}+E_{\rm SC}(L_0)
	 	\nonumber\\
	 	&\equiv\sum_{k=1}^\infty \hbar \omega_k\bar{N}_k^{\beta_A}+E_{\rm SC}(L_0)\, ,
	 \end{align}
	where in the last line we introduced the notation $\bar{N}_k^{\beta_A}$ for the thermal occupation number.
	\item Second, the wall is displaced from $L_0$ to $
		L_1=L(\tau)=L_0(1-\epsilon) $ 
	 following the trajectory (\ref{eq:L}) and compressing the cavity. As a consequence the system evolves under the hamiltonian (\ref{eq:H1}) resulting in the internal energy 
	\begin{equation}
		E_B=\sum_{k=1}^{\infty}\hbar\omega_k(L_1)\text{Tr}({\rho N_k})+E_{\rm SC}(L_1).
	\end{equation}
	
	\item Third, the system is put into contact with a hot bath at inverse temperature $\beta_C$. This leaves the system in a thermal state with internal energy 
	\begin{align}
	E_C=\sum_{k=1}^\infty \hbar \omega_k(L_1)\bar{N}_k^{\beta_C}+E_{\rm SC}(L_1)\, ,
	\end{align}
	where $\bar{N}_k^{\beta_C}$ is also evaluated at $\omega_k(L_1)$.
	\item Fourth, the wall is moved again, returning from $L_1$ to $L_0$,  following the reversed trajectory $\tilde{L}(t)=L(\tau-t) $. That is, the cavity is
		expanded to its original size, ending the process with an internal energy
	\begin{equation}
	E_D=\sum_{k=1}^{\infty}\hbar\omega_k\text{Tr}({\tilde{\rho} N_k})+E_{\rm SC}(L_0),
	\end{equation}
\end{enumerate}
where $\tilde\rho$ is the density matrix associated to the reversed trajectory.

\subsection{Adiabatic evolutions}
If we now assume that the wall is moved slowly enough to be in the condition of the quantum adiabatic theorem, that is if $\tau\gg1/\omega_1$, then there would be no change in the population of the energy levels, meaning
\begin{eqnarray}
	\text{Tr}({\rho N_k})&=&\text{Tr}({\rho^{\beta_A} N_k})\\
	\text{Tr}({\tilde{\rho} N_k})&=&\text{Tr}({\rho^{\beta_C} N_k}).
\end{eqnarray}
Then the heat given to the system would be
\begin{align}
	Q^{\rm Otto}=E_C-E_B=\sum_{k=1}^\infty \hbar \omega_k(L_1)(\bar{N}_k^{\beta_C}-\bar{N}_k^{\beta_A})
\end{align}
and the work 
\begin{align}
W^{\rm Otto}=&(E_A-E_B)+(E_C-E_D)\\\nonumber
=&\sum_{k=1}^\infty \hbar (\omega_k(L_1)-\omega_k)(\bar{N}_k^{\beta_C}-\bar{N}_k^{\beta_A}).
\end{align}
As we can see the static Casimir energy $E_{\rm SC}$ does not modify neither the heat provided nor the work delivered by the machine. Additionally, if the condition $
    \omega_k\beta_A\leq\omega_k(L_1)\beta_C$ 
is satisfied for all $k$, then $Q^{\rm Otto}>0$. Also, using that $L_1<L_0$ we have $\omega_k(L_1)>\omega_k$ and the work ends up being positive $W>0$.

In light of these results, the efficiency can be written as
\begin{align}
\eta^{\rm Otto}&=\frac{W^{\rm Otto}}{Q^{\rm Otto}}\nonumber\\
&=\frac{\sum_{k=1}^\infty \hbar (\omega_k(L_1)-\omega_k)({N}_k^{\beta_C}-\bar{N}_k^{\beta_A})}{\sum_{k=1}^\infty \hbar \omega_k(L_1)(\bar{N}_k^{\beta_C})-\bar{N}_k^{\beta_A}}\nonumber\\
&=\frac{\sum_{k=1}^\infty \hbar \omega_k(L_1)(1-\omega_k/\omega_k(L_1))(\bar{N}_k^{\beta_C}-\bar{N}_k^{\beta_A})}{\sum_{k=1}^\infty \hbar \omega_k(L_1)(\bar{N}_k^{\beta_C}-\bar{N}_k^{\beta_A})}
\end{align}
and, if the spectrum is given by $\omega_k(L)=k\pi/L$ (as is the case in a superconducting circuit choosing the parameters appropriately), we have
\begin{equation}
	\frac{\omega_k}{\omega_k(L_1)}=\frac{k\pi/L_0}{k\pi/L_1}=\frac{L_1}{L_0}=1-\epsilon
\end{equation}
from which we can obtain the following simple result for the efficiency
\begin{equation}
	\eta^{\rm Otto}=\epsilon.
\end{equation}
The above expression implies that the efficiency of the Otto cycle for our system, in the adiabatic limit,  only depends on the compression ratio. In addition, we stress that if we want to achieve the maximum possible efficiency (which is that of Carnot, $\eta^{\rm Carnot}$), then the thermal baths and cavity compression should satisfy
\begin{equation}
	\epsilon=\eta^{\rm Otto}=\eta^{\rm Carnot}=1-\frac{\beta_C}{\beta_A}.
\end{equation}

\subsection{Non-Adiabatic evolutions}
In this Section we will go beyond the adiabatic approximation and show that the DCE induces a sort of ``quantum friction" which reduces the efficiency of the cycle. It is then necessary to calculate
\begin{align}
\text{Tr}({\rho H})=\text{Tr}(U{\rho^{\beta_A}U^\dagger H}).
\end{align}
We shall proceed by using perturbation theory to the lowest  order in $\epsilon$.
Our Hamiltonian Eq.(\ref{eq:H}) is readily suited to perform this calculation in the interaction picture. It is important to remember that in the interaction picture the states evolve according to
\begin{equation}
	|\psi_I(t)\rangle=U_1|\psi_S\rangle
\end{equation}
with $U_1$ given by $$U_1=\mathcal{T}\exp(-i\int_0^t H_1(t)dt/\hbar),$$ while the operators change with time as
\begin{equation}
A_I(t)=U_0A_S(t)U_0^\dagger,
\end{equation}
where $U_0=\exp(-i H_{\text{free}}t/\hbar)$ (we are using the subscript $S$ and $I$ for the Schrödinger and interaction picture respectively).

In our case, the energy of the quantum field at time $t=\tau$ is given by 
\begin{align}
\label{eq:energia1}
E(\tau)&=\text{Tr}({\rho H})=\text{Tr}(U{\rho^{\beta_A}U^\dagger H}) \nonumber\\
&=\text{Tr}({\rho^{\beta_A}U_1^\dagger H_{I}(\tau)U_1}).
\end{align}
We will calculate this energy perturbatively to second order in $\epsilon$. We can do this by first approximating $U_1$ to second order and replacing these results in Eq. (\ref{eq:energia1}) to obtain
\begin{eqnarray}
\label{eq:Napprox}
&&E(\tau)=\text{Tr}(\rho^{\beta_A} H_{I}(\tau))\\
&&-\frac{1}{\hbar}i\int_{0}^{\tau}dt_{1}[\text{Tr}(\rho^{\beta_A} H_IH_{1,I}(t_{1}))-\text{Tr}(\rho^{\beta_A} H_{1,I}(t_{1})H_I)]\nonumber\\
&&+\frac{1}{\hbar^{2}}(-i)^{2}\int_{0}^{\tau}dt_{1}\int_{0}^{\tau}dt_{2}\text{Tr}(\rho^{\beta_A} H_{1,I}(t_{1})H_IH_{1,I}(t_{2}))\nonumber\\
&&+\frac{1}{\hbar^{2}}(-i)^{2}\int_{0}^{\tau}dt_{1}\int_{0}^{t_{1}}dt_{2}\text{Tr}(\rho^{\beta_A} H_{1,I}(t_{2})H_{1,I}(t_{1})H_I)\nonumber\\
&&+\frac{1}{\hbar^{2}}(i)^{2}\int_{0}^{\tau}dt_{1}\int_{0}^{t_{1}}dt_{2}\text{Tr}(\rho^{\beta_A} H_IH_{1,I}(t_{1})H_{1,I}(t_{2})).\nonumber 
\end{eqnarray}
We should further expand the bosonic operators in $\epsilon$, as described in Appendix A,   to obtain 
\begin{align}
H_{I}(\tau)&=\omega(1+\delta(\tau))[N+\frac{\delta(\tau)}{2}(e^{i2\omega\tau}a^{\dagger2}+e^{-i2\omega\tau}a^{2}) \nonumber\\
&+\frac{\delta^{2}(\tau)}{4}(2N+1)]\ .
\label{eq:24}\end{align}

Replacing this result into Eq.(\ref{eq:Napprox}) and further simplifying the expression we find that the internal energy is given by
\begin{eqnarray}
	E(\tau)&=&\sum_{k}\bigg[\hbar\omega_{k}(\tau)\bar{N}_{k}^{\beta_A}\nonumber \\
	&+&\frac{\epsilon^2}{4}\hbar\omega_{k}\int_{0}^{\tau}dt_{1}\int_{0}^{\tau}dt_{2}F^{\beta_A}(t_{1},t_{2})\bigg]
\end{eqnarray}
where, 
\begin{eqnarray}
\label{eq:F}
	&&F^{\beta_A}(t_{1},t_{2})=\frac{\omega_{k}^{\prime2}L_0^2}{\omega_{k}^{2}}\dot{\delta}(t_1)\dot{\delta }(t_2)\cos\left[2\omega_{k}(t_{1}-t_{2})\right] \nonumber \\ &\times &\left\{ 2\bar{N}_{k}^{\beta_A}+1\right\}
	+\sum_{j=0}^{\infty}\dot{\delta}(t_1)\dot{\delta }(t_2)\frac{g_{jk}^{2}}{\omega_{j}\omega_{k}}\nonumber\\
	&&\times\bigg[(\omega_{k}-\omega_{j})^{2}\cos\left[(\omega_{j}+\omega_{k})(t_{1}-t_{2})\right] \nonumber \\ &\times & \left\{ \bar{N}_{k}^{\beta_A}+\bar{N}_{j}^{\beta_A}+1\right\}\nonumber\\ &&+\left(\omega_{j}+\omega_{k}\right)^{2}\cos\left[(\omega_{j}-\omega_{k})(t_{1}-t_{2})\right]\nonumber \\  & \times &\left\{\bar{N}_{j}^{\beta_A}-\bar{N}_{k}^{\beta_A}\right\}\bigg].
\end{eqnarray}
We can see that the non-adiabatic contribution to the energy is
\begin{align}
\label{eq:frictionEnergy}
E_F^{\beta_A}(\tau)=\frac{\epsilon^2}{4}\sum_{k}\hbar\omega_{k}\int_{0}^{\tau}dt_{1}\int_{0}^{\tau}dt_{2}F^{\beta_A}(t_{1},t_{2})\, ,
\end{align}
which is a form of {\it quantum friction} \cite{salamon,abah,stefanatos,stefanatos2,kosloff}. This is because it is non conservative, meaning that it is not a function of the state of the cavity, but rather it fundamentally depends on the trajectory $\delta(t)$ that was used to reach that state. Furthermore, since $E_F$ depends quadratically on $\delta(t)$ and $\dot{\delta}(t)$ this energy contribution will be the same, in modulus and sign, whether the wall moves forward of backwards. This is in direct contrast to conservative forces,  and more alike to the energy dissipated due to a viscous medium that depends on the trajectory (and even the speed $\dot{\delta}$) but not on the direction. 
Furthermore, we can show that this energy is always non-negative. Noting that
\begin{align}
	&\int_0^t dx\int_0^t dy f(x)f(y)\cos(w(x-y)) \\
	&=[\int_0^t dxf(x)\cos(wx)]^2+[\int_0^t dxf(x)\sin(wx)]^2\geq0,\nonumber 
\end{align}
it is clear that the first and second terms in Eq.(\ref{eq:F})
are non-negative. The third term can be written as
\begin{align}
&\sum_{k,j=1}^{\infty}\frac{h_{jk}}{\omega_{j}}\left[\bar{N}_{j}^{\beta_A}-\bar{N}_{k}^{\beta_A}\right]\nonumber\\
&=\sum_{k>j=1}^{\infty}\frac{h_{jk}}{\omega_{j}\omega_{k}}(\omega_{k}-\omega_{j})\left[\bar{N}_{j}^{\beta_A}-\bar{N}_{k}^{\beta_A}\right]\nonumber,
\end{align}
which is non negative because $\omega_{k}>\omega_{j}$ for $k>j$ and therefore $\bar{N}_{j}^{\beta_A}\geq\bar{N}_{k}^{\beta_A}$. This proves that
$E_F^{\beta_A}(\tau)\geq 0$ .

\subsection{An upper bound on the friction energy}

We can set an upper bound on this friction energy by assuming that the wall has vanishing acceleration at the beginning and the end of the motion, $\ddot{\delta}(0)=\ddot{\delta}(\tau)=0$. Integrating by parts
\begin{align}
&	|E_{F}^{\beta_A}(\tau)|=\bigg|\sum_{k=0}^\infty\frac{\hbar\epsilon^{2}}{4}\int_{0}^{\tau}dt_{1}\int_{0}^{\tau}dt_{2}\bigg\{\frac{L_{0}^{2}\omega_{k}^{\prime2}}{16\omega_{k}^5}\dddot{\delta}(t_{1})\dddot{\delta}(t_{2})\nonumber\\
	&\times\cos\left[2\omega_{k}(t_{1}-t_{2})\right]\left\{ 2\bar{N}^{\beta_A}_{k}+1\right\}+\sum_{j=0}^{\infty}\frac{g_{jk}^{2}}{\omega_{j}}\dddot{\delta}(t_{1})\dddot{\delta}(t_{2}) \nonumber\\
	&\times\bigg[\frac{(\omega_{k}-\omega_{j})^{2}}{(\omega_{k}+\omega_{j})^{4}}\cos\left[(\omega_{j}+\omega_{k})(t_{1}^{\prime}-t_{2}^{\prime})\right] \left\{ \bar{N}^{\beta_A}_{k}+\bar{N}^{\beta_A}_{j}+1\right\} \nonumber\\
	&+\frac{(\omega_{k}+\omega_{j})^{2}}{(\omega_{k}-\omega_{j})^{4}}\cos\left[(\omega_{j}-\omega_{k})(t_{1}^{\prime}-t_{2}^{\prime})\right] \left[\bar{N}^{\beta_A}_{j}-\bar{N}^{\beta_A}_{k}\right]\bigg]\bigg\}\nonumber\bigg|\\
	&\leq\left(\int_{0}^{\tau}dt\left|\dddot{\delta}(t)\right|\right)^2\sum_{k=0}^\infty\frac{\hbar\epsilon^{2}}{4}
	\bigg\{\frac{L_{0}^{2}\omega_{k}^{\prime2}}{16\omega_{k}^5} \left( 2\bar{N}^{\beta_A}_{k}+1\right) \nonumber \\ &+\sum_{j=0}^{\infty}\frac{g_{jk}^{2}}{\omega_{j}}
	\bigg[\frac{(\omega_{k}-\omega_{j})^{2}}{(\omega_{k}+\omega_{j})^{4}}\left( \bar{N}^{\beta_A}_{k}+\bar{N}^{\beta_A}_{j}+1\right)\nonumber\\
	&+\frac{(\omega_{k}+\omega_{j})^{2}}{(\omega_{k}-\omega_{j})^{4}}\left(\bar{N}^{\beta_A}_{j}-\bar{N}^{\beta_A}_{k}\right)\bigg]\bigg\}.
\end{align}
If we further assume that the acceleration of the wall has one local (and global) maximum at $t_M$ and one negative local (and global) minimum at $t_m$ then
\begin{align}
\label{eq:bound}
|E_{F}^{\beta_A}(\tau)|&\leq\left(\ddot{\delta}(t_M)-\ddot{\delta}(t_m)\right)^2\sum_{k=0}^\infty\hbar\epsilon^{2} \nonumber\\
&\times\bigg\{\frac{L_{0}^{2}\omega_{k}^{\prime2}}{16\omega_{k}^5} \left( 2\bar{N}^{\beta_A}_{k}+1\right)+\sum_{j=0}^{\infty}\frac{g_{jk}^{2}}{\omega_{j}}\nonumber\\ &\bigg[\frac{(\omega_{k}-\omega_{j})^{2}}{(\omega_{k}+\omega_{j})^{4}}\left( \bar{N}^{\beta_A}_{k}+\bar{N}^{\beta_A}_{j}+1\right)\nonumber\\
&+\frac{(\omega_{k}+\omega_{j})^{2}}{(\omega_{k}-\omega_{j})^{4}}\left(\bar{N}^{\beta_A}_{j}-\bar{N}^{\beta_A}_{k}\right)\bigg]\bigg\}.
\end{align}
In this way, by making a few reasonable assumptions, we can bound the friction energy due to the DCE over a very wide range of trajectories. 

\subsection{Efficiency as a heat engine}

From the energy generated by the moving wall, it is straightforward to calculate the heat dissipated
\begin{align}
Q= &E_{C}-E_{B}=\sum_{k=1}^{\infty}\hbar\omega_{k}(L_{1})\bar{N}_{k}^{\beta_{C}}-\sum_{k=1}^\infty\bigg[\hbar\omega_{k}(L_{1})\bar{N}_{k}^{\beta_{A}}\nonumber\\
&+\epsilon^2\frac{\hbar\omega_{k}(L_1)}{4}\int_{0}^{\tau}dt_{1}\int_{0}^{\tau}dt_{2}F^{\beta_{A}}(t_{1},t_{2})\bigg].
\end{align}
Since the friction energy is already quadratic in $\epsilon$, we can approximate $\omega_{k}(L_{1})\approx\omega_{k}$ in that term. The heat is then given by
\begin{align}
Q&=Q^{\rm Otto}-E_{F}^{\beta_{C}}(\tau),
\end{align}
to second order in $\epsilon$.

The work is also easy to calculate
\begin{align}
	W&=W_{AB}+W_{CD}\nonumber\\
	&=W^{\rm Otto}-\left[E_{F}^{\beta_{A}}+E_{F}^{\beta_{C}}\right].
\end{align}
Finally, the efficiency as a heat engine for the non-adiabatic cycle is given by
\begin{align}
	\eta&=\frac{W^{\rm Otto}-\left[E_{F}^{\beta_{A}}+E_{F}^{\beta_{C}}\right]}{Q^{\rm Otto}-E_{F}^{\beta_{C}}}\nonumber\\
	&\approx\frac{W^{\rm Otto}}{Q^{\rm Otto}}+\frac{W^{\rm Otto}E_{F}^{\beta_{C}}-Q^{\rm Otto}\left[E_{F}^{\beta_{A}}+E_{F}^{\beta_{C}}\right]}{(Q^{\rm Otto})^{2}}+\mathcal{O}(\epsilon^{3})\nonumber\\
	&\approx\eta^{\rm Otto}-\frac{E_{F}^{\beta_{A}}+E_{F}^{\beta_{C}}}{Q^{\rm Otto}}+\mathcal{O}(\epsilon^{3}),
\end{align}
which is always smaller than $\eta^{\rm Otto}$.
Futhermore, we can use Eq.(\ref{eq:bound}) to set a lower bound on the efficiency of the Otto cycle on the non-adiabatic regime:
\begin{align}
&\eta  \geq\eta^{Otto}-\frac{|E_{F}^{\beta_{A}}+E_{F}^{\beta_{C}}|}{Q^{Otto}}\nonumber\\
&\geq\eta^{Otto} \nonumber \\ &- \frac{\left(\ddot{\delta}(t_M)-\ddot{\delta}(t_m)\right)^2}{Q^{otto}}\sum_{k=0}^\infty\hbar\epsilon^{2}
\bigg\{\frac{L_{0}^{2}\omega_{k}^{\prime2}}{16\omega_{k}^5} \left( 2\bar{N}^{\beta_A}_{k}+1\right)\nonumber \\ &+\sum_{j=0}^{\infty}\frac{g_{jk}^{2}}{\omega_{j}}\bigg[\frac{(\omega_{k}-\omega_{j})^{2}}{(\omega_{k}+\omega_{j})^{4}}\left( \bar{N}^{\beta_A}_{k}+\bar{N}^{\beta_A}_{j}+1\right)\nonumber \\
&+\frac{(\omega_{k}+\omega_{j})^{2}}{(\omega_{k}-\omega_{j})^{4}}\left(\bar{N}^{\beta_A}_{j}-\bar{N}^{\beta_A}_{k}\right)\bigg]\nonumber\\
&+\frac{L_{0}^{2}\omega_{k}^{\prime2}}{16\omega_{k}^5} \left( 2\bar{N}^{\beta_C}_{k}+1\right)+\sum_{j=0}^{\infty}\frac{g_{jk}^{2}}{\omega_{j}}\nonumber\\ &\bigg[\frac{(\omega_{k}-\omega_{j})^{2}}{(\omega_{k}+\omega_{j})^{4}}\left( \bar{N}^{\beta_C}_{k}+\bar{N}^{\beta_C}_{j}+1\right)\nonumber\\
&+\frac{(\omega_{k}+\omega_{j})^{2}}{(\omega_{k}-\omega_{j})^{4}}\left(\bar{N}^{\beta_C}_{j}-\bar{N}^{\beta_C}_{k}\right)\bigg]\bigg\} .
\end{align}

\section{An example}\label{sec:tray}
In order to illustrate our results,  we consider the trajectory given by the lowest order polynomial 
\begin{align}
\label{eq:poli}
\delta(t)=10(t/\tau)^3-15(t/\tau)^4+6(t/\tau)^5.
\end{align}
satisfying all the following conditions  
\begin{align}
\label{derivatives}
	\delta(0)&=\dot{\delta}(0)=\ddot{\delta}(0)=0, \\
	\dot{\delta}(\tau)&=\ddot{\delta}(\tau)=0, \nonumber\\
	\delta(\tau)&=1.\nonumber
\end{align}

\begin{figure}
	\begin{center}
		\includegraphics[scale=0.6]{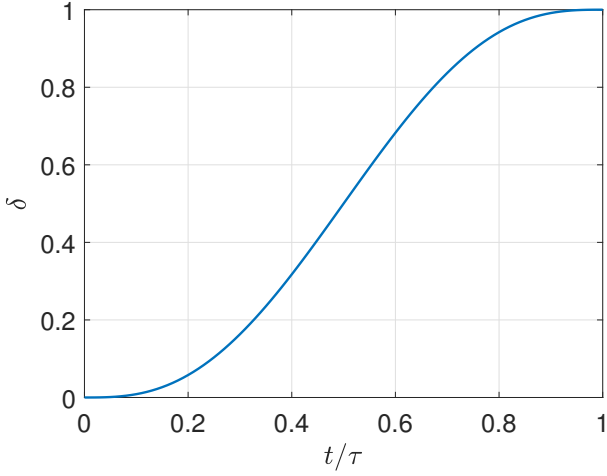}
		\caption{Trajectory chosen to exemplify our results, verifying the constraints imposed in the derivatives of Eq.(\ref{derivatives}).}
		\label{fig:tray}
	\end{center}
\end{figure}
\begin{figure}
	\begin{center}
		\includegraphics[scale=0.6]{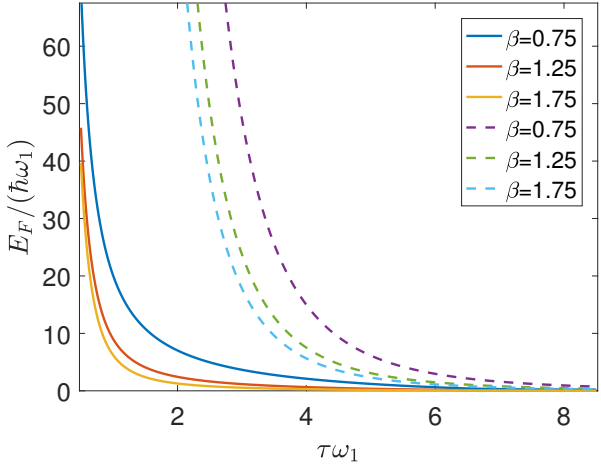}
		\caption{Friction energy as a function of dimensionless time, for different values of $\beta$. In solid lines, we present the friction energy $E_F$ computed in each case, while in dashed lines, we plot the corresponding upper bound derived. The compression ratio used in all cases is $\epsilon=0.01$.}
		\label{fig:E_t}
	\end{center}
\end{figure}
\begin{figure}
	\begin{center}
		\includegraphics[scale=0.6]{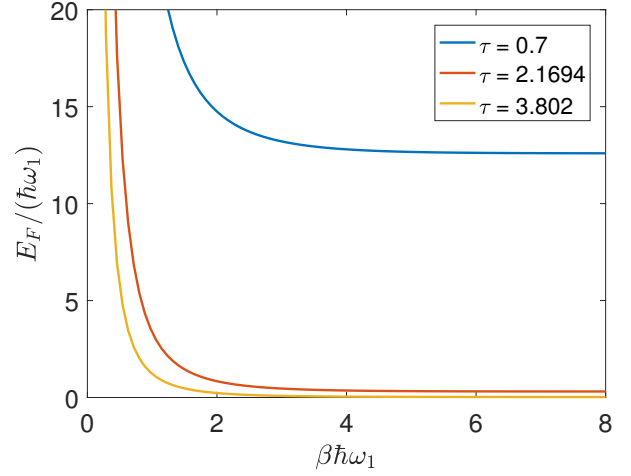}
		\caption{Friction energy as a function of $\beta \hbar \omega_1$, for different values of $\tau$. Photon production increases as the time period $\tau$  decreases, and therefore the friction energy becomes more important. The compression ratio used is $\epsilon=0.01$.}
		\label{fig:E_beta}
	\end{center}
\end{figure}

This trajectory is shown in Fig. (\ref{fig:tray}). In Fig. (\ref{fig:E_t}), we present the corresponding friction energy produced by the DCE. It is important to note that it vanishes for slow motions ($\tau\to\infty$) and becomes arbitrarily large for sudden movements of the wall. On the other hand, for a fixed $\tau$, it converges to a finite value as the inverse temperature of the initial state $\beta$ grows (Fig. (\ref{fig:E_beta})). This is clearly associated to the photon production given by the DCE which arises even from a vacuum state.

\begin{figure}
	\begin{center}
		\includegraphics[scale=0.6]{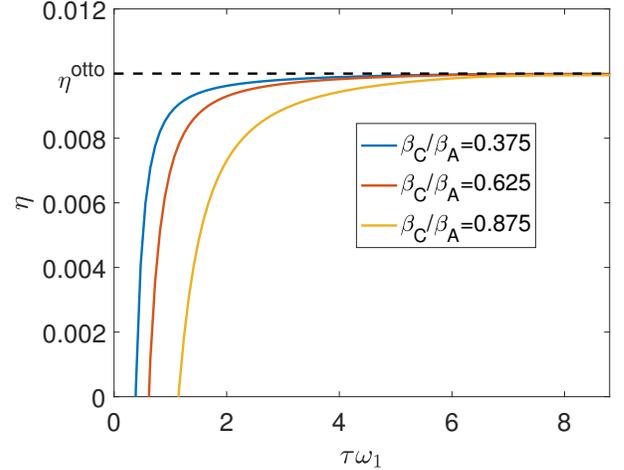}
		\caption{ Efficiency of the engine for different relations among the temperature of the baths ($\beta_C/\beta_A$) as a function of $\tau\omega_1$. The quantum engine operated in the non adiabatic regime has a smaller efficiency compared to the adiabatic case, indicated as $\eta^{\rm Otto}$ in the plot. The compression ratio used is $\epsilon=0.01$.}
		\label{fig:eta}
	\end{center}
\end{figure}

As we have mentioned before, the friction energy of the DCE is always non-negative and, just as the friction on a classical piston, it  diminishes the efficiency of the engine from the Otto eficiency  $\eta^{\rm Otto}$.
In general the quantum friction increases as the motion of the wall becomes more sudden ($\tau\to0$). In fact, because of this there is a minimum timescale $\tau$ below which the engine can no longer function as such, since the work produced becomes negative (see Fig. \ref{fig:eta}).

\begin{figure}
	\begin{center}
		\includegraphics[scale=0.6]{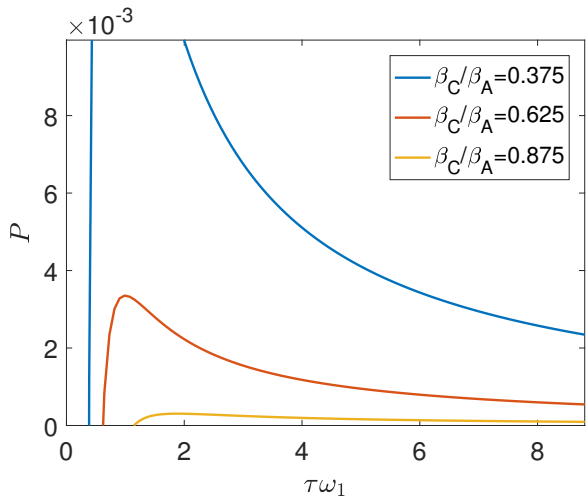}
		\caption{ Power produced by the engine for different temperature ratios as a function of time. For each ratio it has a peak at approximately $\tau\omega_1\sim1$ indicating an optimal timescale of operation. The compression ratio used was $\epsilon=0.01$.}.
		\label{fig:P}
	\end{center}
\end{figure}

We further study the power $P$  produced by this engine. In the adiabatic case,  when $\tau\omega_1\gg 1$, the work is independent of the time scale $\tau$ and thus the power increases as the time decreases,  since $P\sim 1/\tau$. In the (non-adiabatic) opposite
limit $\tau\omega_1\ll 1$, the power turns out to be proportional to $1/\tau^4$. Then the power will have two contributions, with different signs, that scale with different powers of $\tau$. 
As a consequence, we expect a peak around the time the friction energy becomes relevant, $\tau\omega_1\sim1$. This is illustrated 
in Fig. \ref{fig:P}.

\begin{figure}
	\begin{center}
		\includegraphics[scale=0.6]{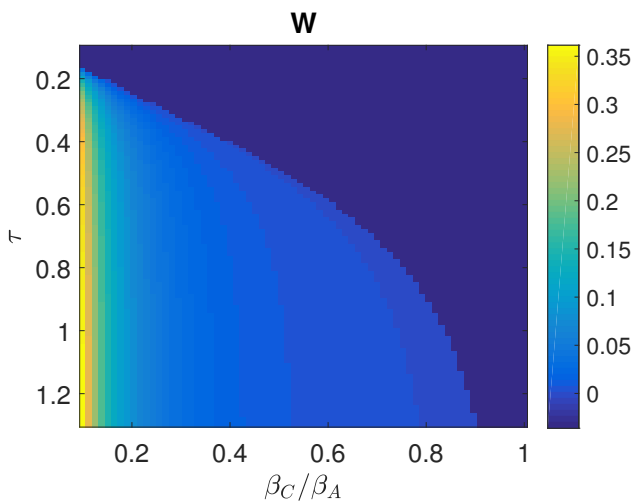}
		\caption{ Work extracted from the engine. We can identify regions where no work can be extracted from the Otto cycle.  The compression ratio used is $\epsilon=0.01$.}
		\label{fig:w}
	\end{center}
\end{figure}

Finally,  in Fig. (\ref{fig:w}) we represent the extracted work.
Therein, we can note that $W>0$ for large ratios of the bath temperatures ($\beta_C/\beta_A$) and longer times $\tau$. Otherwise, if $\tau\rightarrow 0$ and $\beta_C/\beta_A\rightarrow 1$, we note that the work vanishes.

\section{Superconducting circuit refrigerator}\label{sec:implementation}
It is also possible to implement a quantum field refrigerator using this system, that is,  a quantum system that cools a cold bath and heats a hot bath while consuming work. In this section, we shall compute the coefficient of performance of the
Otto refrigerator and the effect of quantum friction on it. 

\subsection{Adiabatic evolutions} 
In the adiabatic case the heat taken from the cold reservoir is given by 
\begin{align}
	Q^{\rm Otto}&=E_{A}-E_{D}\nonumber\\
	&=\sum_{k}\hbar\omega_{k}\left[N_{k}^{\beta_{A}}(\omega_{k})-N_{k}^{\beta_{C}}(\omega_{k}(L_{1}))\right]
\end{align}
which is positive as long as
\begin{equation}
	\frac{L_{1}}{L_{0}}\leq\frac{\beta_{C}}{\beta_{A}}\leq1.
\end{equation}
These conditions  imply that the work consumed by the system 
\begin{align}
	W^{\rm Otto}&=(E_{B}-E_{A})+(E_{D}-E_{C})\\
	&=\sum_{k}\hbar\left[\omega_{k}(L_{1})-\omega_{k}\right]\left[N_{k}^{\beta_{A}}(\omega_{k})-N_{k}^{\beta_{C}}\omega_{k}(L_{1})\right], \nonumber
\end{align}
is also positive. We measure the efficiency of this machine by the coefficient of performance
\begin{align}
	\eta^{\rm Otto}&=\frac{Q^{\rm Otto}}{W^{\rm Otto}}
	&=\frac{1}{\omega_{k}(L_{1})/\omega_{k}-1}
	=\frac{1}{\epsilon}-1,
\end{align} 
assuming an equidistant spectrum and $L_1=L_0(1-\epsilon)$.

\subsection{Non-adiabatic case} 
\begin{figure}
	\begin{center}
		\includegraphics[scale=0.6]{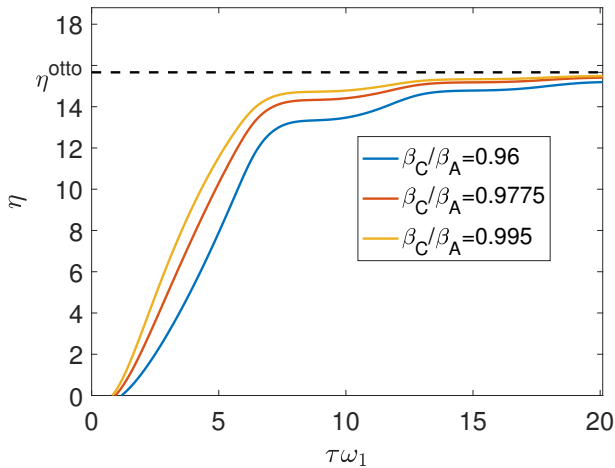}
		\caption{ The coefficient of performance at finite time is always lower than the one for the adiabatic regime. The compression ratio used was $\epsilon=0.06$.}
		\label{fig:eta_heladera}
	\end{center}
\end{figure}

On the other hand, if the motion of the wall is non adiabatic it is necessary to include the friction energy due to the DCE. 
As we have previously proceeded for the heat machine, we can compute the heat extracted by the refrigerator. In that case,  the coefficient of performance in the non-adiabatic regime is given by
\begin{align}
\eta=\frac{Q^{\rm Otto}-E_{F}^{\beta_{C}}}{W^{\rm Otto}+E_{F}^{\beta_{A}}+E_{F}^{\beta_{C}}}\leq\frac{Q^{\rm Otto}}{W^{\rm Otto}}
\end{align} 
which is always lower than the adiabatic one. In Fig.(\ref{fig:eta_heladera}), we show the coefficient of performance at finite time for different rates of bath temperatures $\beta_C/\beta_A$. It can easily be noted that the coefficient of performance is lower than the corresponding adiabatic one ($\eta^{\rm Otto}$). We can also note a hierarchy in the rate of the bath temperature: the more similar the thermal baths are, the bigger the coefficient of performance we can achieve.

\begin{figure}
	\begin{center}
		\includegraphics[scale=0.6]{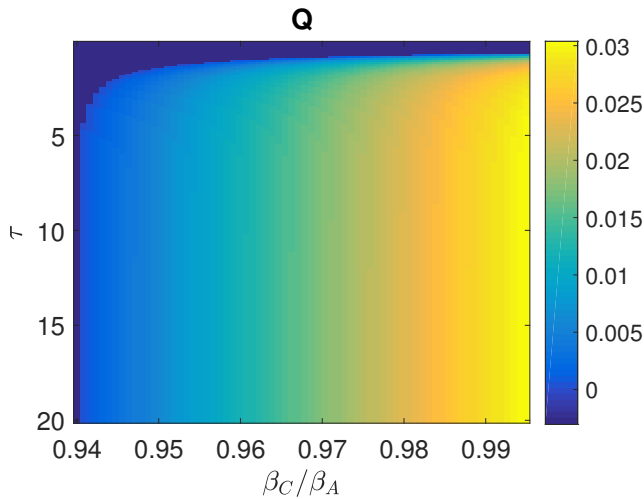}
		\caption{ The heat extracted by the refrigerator $Q$ is greater for baths with similar temperatures and a large timescale of operation $\tau$, otherwise it can even be negative, meaning the machine stops working as a refrigerator. The compression ratio used was $\epsilon=0.06$.}
		\label{fig:heat_heladera}
	\end{center}
\end{figure}

In Fig. (\ref{fig:heat_heladera}) we present the heat extracted $Q$ by the refrigerator for different rates of thermal temperatures $\beta_C/\beta_A$ and different operation timescales $\tau$. We can note that there are small values of $\tau$ for which the heat $Q$ becomes negative, implying that the machine is no longer working as a refrigerator. The greatest values of $Q$ extracted are for similar bath temperatures and longer operational timescales.

\begin{figure}
	\begin{center}
		\includegraphics[scale=0.6]{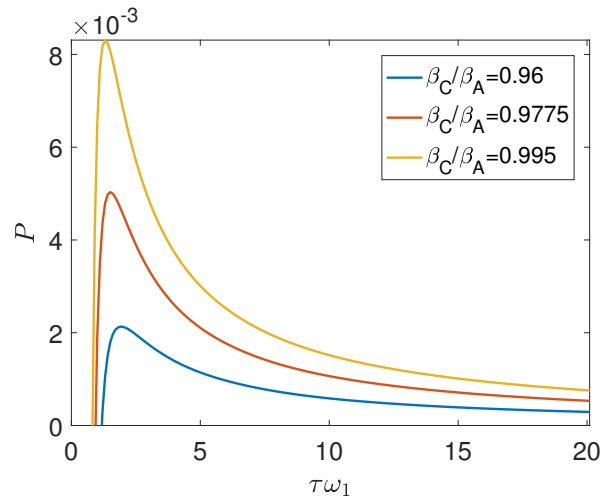}
		\caption{ The cooling power of the refrigerator also has a peak at an optimal timescale of $\tau\omega_1\sim 1$. The compression ratio used was $\epsilon=0.06$.}
		\label{fig:P_heladera}
	\end{center}
\end{figure}

Finally in Fig.(\ref{fig:P_heladera}), we show the cooling power for the refrigerator for growing operational timescales $\tau$. We can note that there is a maximum value for the cooling power when $\tau \omega_1 \sim 1$, achieving a greater value when the thermal bath have similar temperatures $\beta_C/\beta_A \rightarrow 1$.

\section{Avoiding quantum friction from the DCE}
As we have seen in the previous sections,  the motion of the wall produces photons from the vacuum through the DCE,  which causes the efficiency of the cycle to be reduced from the adiabatic case. However if we assume that the spectrum of the cavity is equidistant,  $\omega_n=n\pi/L_0$, there is a class of trajectories for which the friction energy vanishes.  Indeed, let us assume that the trajectory of the wall is such that
\begin{equation}
    {\delta}(t)=G(t+L_0)-G(t-L_0),
\end{equation}
where $G(x)$ is a smooth function that is linear for $x<0$ and $x>\tau$.  Then the motion starts at $t=-L_0$, ends at $t=L_0+\tau$, and one can show that the friction energy vanishes at order $\epsilon^2$. Indeed, we can see this by checking explicitly that,  for an arbitrary function $G$, the integrals that appear in Eq. \eqref{eq:Napprox}
\begin{align}
    I_n(t)&=\int_{-L_0}^{t}(\dot G(t^\prime+L_0)-\dot G(t^\prime-L_0))\cos(n\pi t^\prime/L_0)dt^\prime\nonumber\\
    J_n(t)&=\int_{-L_0}^{t}(\dot G(t^\prime+L_0)-\dot G(t^\prime-L_0))\sin(n\pi t^\prime/L_0)dt^\prime
\end{align}
do vanish at $t=L_0+\tau$. Therefore, the friction energy in Eq. \eqref{eq:frictionEnergy} also vanishes at order $\epsilon^2$. 
\begin{figure}
	\begin{center}
		\includegraphics[scale=0.6,trim={1.5cm 0 0 0},clip]{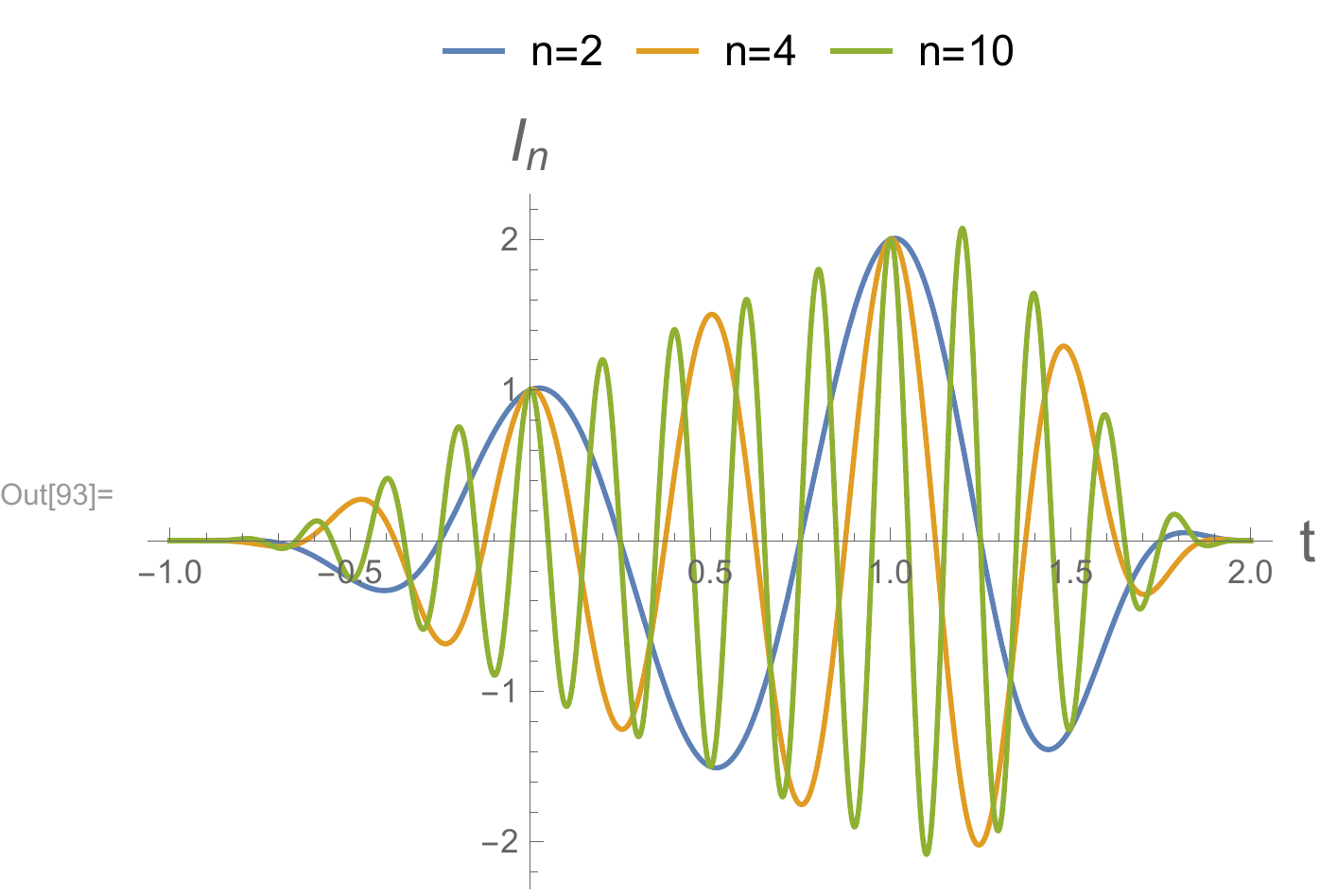}
		\caption{ Integrals $I_n$ for $n=2,4,10$ as function of time for the case where the function $\dot{G}(x)$ is given by the polynomial in Eq. (\ref{eq:poli}). The compression ratio used was $\epsilon=0.7$, $\tau=1$, $L_0=1$}
		\label{fig:integrales_atajo}
	\end{center}
\end{figure}

Some examples of these integrals are shown in Fig. \ref{fig:integrales_atajo}, 
for the case where the function $\dot{G}(x)$ vanishes for $x<0$, is given by the polynomial in Eq. (\ref{eq:poli}) for $0<x<\tau$ and is constantly 1 for $x>\tau$. The plots illustrate the physical mechanism behind the avoidance of
the friction coming from the DCE: even though photons are initially generated as the wall moves,  they are then reabsorbed at a later time, leaving the system with the populations in each mode unchanged.


\section{Conclusions}\label{sec:conc}

We have discussed several aspects of the Otto cycle in a superconducting cavity, modeled with a quantum scalar field confined to a one dimensional cavity with a moving boundary. In particular, we have considered the adiabatic approximation and found the efficiency for the cycle when the machine operates as a heat engine,  and the coefficient of performance when it operates as a refrigerator. Furthermore, we have solved the time evolution of the field up to second order in the compression ratio,  and used it to evaluate the efficiency of the cycle in finite time. We have seen that the motion of the wall generates photons through the DCE, which are associated with a  form of quantum friction energy. We have shown that this energy is non-negative,  and we have been able to give an upper bound for it. These results were then used to calculate the efficiency of the engine and show that a finite time operation leads to a reduced efficiency as compared with the adiabatic case. This is due to the energy wasted in producing the dynamical Casimir photons. We have also calculated the efficiency and power for a typical trajectory of the wall, finding that the friction energy grows very rapidly as the timescale $\tau$ of the motion is reduced. This fact in turn leads to a peak in the power output of the machine and, eventually, for a small $\tau$ causes the machine to stop working as an engine. We have also studied the use of the cycle for a quantum field refrigerator operated in finite time finding similar results: a loss of efficiency and bounded cooling power.

Finally, we have used the explicit expression found for the friction energy to show that there exists a family of trajectories of the wall for which the friction vanishes (at least to second order in the compression ratio). Physically this can be understood as a photon generation via the DCE,  followed by a re-absortion of the photons by the wall at a later time. These trajectories would be  extremely useful to eliminate the losses and improve the efficiency of the machine. 



\appendix
\section{The Hamiltonian}\label{apen1}

In this Appendix we will show that the Hamiltonian for a scalar quantum field with a moving wall can be described by Eq.(\ref{eq:H1}).
We start with the Lagrangian for the scalar field

\begin{equation}
L=\frac{1}{2}\int_{0}^{L_{0}}dx\left[\left(\partial_{t}\Phi\right)^{2}-\left(\partial_{x}\Phi\right)^{2}\right]
\end{equation}
and expand the field in an instantaneous basis for each $L$ as
\begin{equation}
\Phi=\sum_{n=1}^{\infty}Q_{n}(t)\psi_{n}(x)\, ,
\end{equation}
where
\begin{equation}
\psi_{n}(x)=\sqrt{\frac{2}{L_{0}}}\sin(\omega_{n}(L)x).
\end{equation}

Introducing the definitions and relations
\begin{align}
&\int_{0}^{L}dx\psi_{k}\psi_{j}=\delta_{kj},\nonumber\\
&L\int_{0}^{L}dx\left(\partial_{L}\psi_{k}\right)\psi_{j}=g_{kj}=-g_{jk},\nonumber\\
&\sum_{l=1}^{\infty}g_{lk}g_{lj}=L^{2}\int_{0}^{L}\partial_{L}\psi_{k}\partial_{L}\psi_{j},\nonumber\\
&\int_{0}^{L}dx\psi_{k}^{\prime}\psi_{j}^{\prime}=\omega_{k}^{2}(L)\delta_{kj},
\end{align}
the Lagrangian reads, to first order in $\dot{L}/L$:
\begin{equation}
\mathcal{L}=\frac{1}{2}\sum_{k=1}^{\infty}\big[\dot{Q}_{k}^{2}-\omega_{k}^{2}(L)Q_{k}^{2}\big]+\frac{1}{2}\sum_{k,j=1}^{\infty}\frac{\dot{L}}{L}g_{kj}\big[\dot{Q}_{k}Q_{j}-Q_{k}\dot{Q}_{j}\big].
\end{equation}
We now turn to the Hamiltonian formulation. The canonical momentum is given by
\begin{equation}
P_{i}=\frac{\partial\mathcal{L}}{\partial\dot{Q}_{i}}=\dot{Q}_{i},
\end{equation}
from which we obtain the Hamiltonian
\begin{align}
H&=\sum_{k=1}^{\infty}\dot{Q}_{k}P_{k}-\mathcal{L}\nonumber\\
&=\frac{1}{2}\sum_{k=1}^{\infty}\big[P_{k}^{2}+\omega_{k}^{2}(L)Q_{k}^{2}\big]-\sum_{k,j=1}^{\infty}\frac{\dot{L}}{L}g_{kj}P_{k}Q_{j}.
\end{align}
We now quantize the theory by promoting the coordinates and momenta to operators with the canonical commutation relations
\begin{equation}
Q_{k}\to\hat{Q}_{k},\quad P_{k}\to\hat{P}_{k},\quad[\hat{Q}_{k},\hat{P}_{k}]=i
\end{equation}
and define the bosonic operators
\begin{align}
a_{k1}&:=\frac{1}{\sqrt{2\omega_{k}(L)}}(\omega_{k}(L)\hat{Q}_{k}+i\hat{P}_{k})\nonumber\\
a_{k1}^{\dagger}&:=\frac{1}{\sqrt{2\omega_{k}(L)}}(\omega_{k}(L)\hat{Q}_{k}-i\hat{P}_{k}).
\end{align}
Since we are working to first order in $\delta L/L$,  we approximate
\begin{equation}
a_{k1}\approx a_{k}+\delta L\frac{\omega_{k}^{\prime}}{2\omega_{k}}a_{k}^{\dagger}.
\end{equation}
Then, replacing in the Hamiltonian, we get
\begin{align}
H&=\sum_{k=1}^{\infty}\left[\omega_{k}a_{k}^{\dagger}a_{k}+\omega_{k}^{\prime}\delta La_{k}^{\dagger}a_{k}+\delta L\frac{\omega_{k}^{\prime}}{2}\left(a_{k}^{2}+a_{k}^{\dagger2}\right)\right]\nonumber\\
&+\frac{1}{2i}\sum_{k,j=1}^{\infty}\frac{\dot{L}}{L_{0}}g_{kj}\sqrt{\frac{\omega_{k}}{\omega_{j}}}\big[a_{k}a_{j}-a_{j}^{\dagger}a_{k}+a_{j}a_{k}^{\dagger}-a_{k}^{\dagger}a_{j}^{\dagger}\big].
\end{align}

\section*{Acknowledgements}
This work was supported by ANPCyT, CONICET, Universidad de Buenos Aires and Universidad de Cuyo, Argentina. 

\end{document}